# Mass and generalized Thiele equation of the magnetic skyrmion

Yoshishige Suzuki[1, 2, †], Soma Miki[1,2], Hikaru Nomura[1,2], and Eiiti Tamura[1,2]

[1]Graduate School of Engineering Science, Osaka University, Toyonaka, Osaka 560-8531, Japan

[2]Spintronics Research Network Division, Institute for Open and Transdisciplinary Research Initiatives, Osaka University, Yamadaoka 2-1, Suita, Osaka, 565-0871, Japan

Abstract

An analytical expression is obtained for the mass of an isolated magnetic skyrmion and its linearized equation of motion. The magnetic skyrmion is viewed as a topologically protected spin-wave soliton in the magnetic ultrathin films stabilized by the interfacial-Dzyaloshinskii-Moriya interaction. The equations of motion are derived from the Landau-Lifshitz-Gilbert equation for both the skyrmion charge and magnetization centers. They are generalized Thiele equations, including the gyro-term, dissipation term, external force, acceleration term with the tensorial mass, and time derivatives of the external forces. The equation of motion of the center of the skyrmion charge essentially shows the massless nature of the skyrmion. In contrast, the equation of motion for the magnetization center results in a finite mass that is in the same order as the Döring mass density for the linear domain wall. Furthermore, the time derivative of the external force predominantly contributes to the immediate response of the skyrmion motion, i.e., the mass-less property remains even after the skyrmion acquires its kinetic mass.

A micromagnetic simulation based on the LLG equation was performed for various magnetic parameters. Obtained trajectories at 0 K are compared with the theoretical predictions.

---

[†]Corresponding author: Yoshishige SUZUKI
 e-mail: suzuki.yoshishige.es@osaka-u.ac.jp

**Introduction**

Skyrmion was introduced in 1962 by Skyrme to explain the stability of the elementary particles from the topological nature of the vector field [1]. For the 2-dimensional magnetic texture, the skyrmion characteristics are defined by projecting the 2-dimensional plane $\mathbb{R}^2$ onto the 2-dimensional sphere $S^2$ [2]. An isolated skyrmion is a magnetic soliton exciting a local magnetization reverse for the host magnetic film showing a perpendicular magnetization and possessing a non-zero integer skyrmion number (winding number/skyrmion charge) [3].

$$n_{Sk} = \frac{1}{4\pi} \int_{-\infty}^{+\infty} dxdy\, \mathbf{m} \cdot \left( \frac{\partial \mathbf{m}}{\partial x} \times \frac{\partial \mathbf{m}}{\partial y} \right), \qquad (1)$$

where the 3-dimensional vector of unit modulus $\mathbf{m}$ is the magnetization direction that depends on the 2-dimensional *x-y* coordinates.

In 2006, the existence of stable magnetic skyrmion was predicted theoretically [4] and soon observed in a natural material using neutron scattering [5]. The first observed skyrmion is a cylindrical Bloch-skyrmion, which is stabilized by bulk-type Dzyaloshinskii-Moriya interaction (DMI). After a while, disk shape skyrmion is discovered in the ultrathin magnetic metallic film [6]. The interfacial DMI [7] stabilizes Néel-type skyrmion in the ultrathin film because of a lack of inversion symmetry at the interface. The diameter of the skyrmion varies from a few nanometers to several micrometers.

The skyrmion is the smallest magnetic domain and can be controlled by a small electric current [8, 9,10] and/or a voltage [11,12]. In addition, it shows peculiar diffusion called "gyro-diffusion" [13, 14]. Therefore, it is an interesting object for fundamental physics and applications.

Before the prediction of the magnetic skyrmion, the kinetics in topological magnetic structures, like domain wall, vortex, and the Bloch point, attracted attention [15]. In 1973, the Thiele equation [16], which describes the steady-state motion of magnetic structures, was derived from the Landau-Lifshitz-Gilbert equation [17,18].

$$-G_{xy}\mathbf{e}_z \times \dot{\mathbf{X}} - \Gamma_{xx}\dot{\mathbf{X}} + \mathbf{F} = 0 \qquad (2)$$

$$\begin{cases} G_{xy} = 4\pi\, a\, n_{Sk} \\ \Gamma_{xx} = \alpha\, a \int_{-\infty}^{+\infty} dxdy\, \frac{\partial \mathbf{m}}{\partial x} \cdot \frac{\partial \mathbf{m}}{\partial x} \\ a \equiv \frac{\mu_0 d M_s}{-\gamma} \end{cases} \quad (3)$$

In the above equations, $\dot{\mathbf{X}} = \frac{d}{dt}\left(X_x(t), X_y(t)\right)$ is a time derivative of the 2-dimensional position of the object (velocity), and $\mathbf{F}$ the force, $G_{xy}$ the size of the gyro-vector, $\Gamma_{xx}$ the dissipation dyadic, $\alpha$ the Gilbert damping constant, $M_s$ the saturation magnetization, $d$ the film thickness, $\mu_0 = 4\pi \times 10^{-7}$ H/m is the vacuum magnetic permeability, $\gamma < 0$ the gyro-magnetic ratio, and $a$ is the angular momentum density. As $G_{xy}$ is proportional to the skyrmion number, one may find that the dynamics of the magnetic structure are governed by the topological nature of the magnetic structure.

The Thiele equation is correct as far as steady-state motion is considered. In the case of transient dynamics, however, the magnetic structure changes its form, and the Thiele equation is not applicable anymore. In order to describe the transient dynamics of the magnetic domain wall, the Döring mass was introduced in the 1940s [19]. Accordingly, for the skyrmion, an acceleration term with finite mass was introduced to the Thiele equation [20]. Then an experimental observation was done using the X-ray microscope [21]. However, the observed mass was a hundred times larger than the theoretical estimation. During the last decade, many papers concerning the skyrmion mass have appeared. Among papers, the frequency dependence of the mass [22], frequency-independent mass [14], zero-mass of the bounded spin waves [23], quantum contribution [24], *etc*.[25, 26] have been discussed.

In this paper, to understand the fundamental property of the intrinsic mass of skyrmions, we derive a linearized equation of motion of an isolated Néel skyrmion in magnetic ultrathin films from the LLG equation. All intrinsic interactions except the interaction with extended spin waves are included in the calculation. The theory treats small skyrmions [27] and relatively large skyrmion bubbles, which are stabilized by magnetic dipole interaction [28, 29]. We proceed with our derivation as follows. First, the equation of motion of the collective coordinates [30] was block-diagonalized concerning the symmetry of the system within linear response approximation. Then the contribution of an additional degree of freedom was erased to construct a closed expression of the equation of motion of the skyrmion position. The obtained equation of motion includes the 2×2 tensorial mass and the time derivative of the external forces. Finally, the results are quantitatively compared with micromagnetic simulation based on the LLG equation.

**LLG equation and collective coordinates**

In this paper, disk shape skyrmion with $n_{Sk} = \pm 1$ will be treated (see Fig. 1 (a)). The topological structure of the skyrmion is determined by vorticity $v$, helicity $h$, polarization $p$, and skyrmion number $n_{Sk}$ [3]. The magnetization-direction vector $\mathbf{m}$ in equilibrium, at a position $(r\cos\varphi, r\sin\varphi)$ on the x-y plane, relates to the topological numbers as follows;

$$\begin{cases} \mathbf{m}(r,\varphi) = \begin{pmatrix} \sin\theta\cos\phi \\ \sin\theta\sin\phi \\ \cos\theta \end{pmatrix} \\ \theta = \theta(r) \\ \phi = v\varphi + h \\ \cos\theta|_{r=0} - \cos\theta|_{r=\infty} = 2p \end{cases} \quad (4)$$

The vorticity $v = -1$ corresponds to the anti-skyrmion, and the case with $v = +1$ (skyrmion) is the subject of this paper. The skyrmion number is determined by the rotations of $\theta(r)$ when $r$ changes from 0 to infinite. The half rotation gives $n_{Sk} = \pm 1$. The z-component of the magnetization direction vector at the center of the skyrmion is the polarity of the skyrmion. The helicity $h = 0, \pi$ corresponds to the Néel skyrmion and $h = \pm \pi/2$ the typical Bloch skyrmion. If the interfacial DMI, bulk DMI, or dipole interaction coexist, the helicity may take an intermediate value [31]. In this paper, we also refer to those skyrmions with intermediate helicity at equilibrium as Bloch skyrmions.

Those states are ground or metastable states of the internal energy functional $U^{int}[\mathbf{m}]$. The internal energy consists of exchange energy $U^{ex}$, interfacial DMI energy $U^{DMI}$, magnetic anisotropy energy $U^{ani}$, Zeeman energy $U^{Zeeman}$, and dipole coupling energy $U^{dipole}$ [31] (see Appendix I). The internal energy is expressed by the aforementioned material parameters, and the magnetization distribution $\mathbf{m}(\mathbf{x})$ as follows;

$$U^{int} = U^{ex} + U^{DMI} + U^{ani} + U^{Zeeman} + U^{dipole}$$

$$= \int d^3x \left( \begin{array}{l} A_{ex}|\nabla_{\mathbf{x}}\mathbf{m}(\mathbf{x})|^2 + D\big(m_z(\mathbf{x})(\vec{\nabla}_{\mathbf{x}} \cdot \mathbf{m}(\mathbf{x})) - (\mathbf{m}(\mathbf{x}) \cdot \vec{\nabla}_{\mathbf{x}})m_z(\mathbf{x})\big) + K_u\big(1 - (\mathbf{e}_z \cdot \mathbf{m}(\mathbf{x}))^2\big) \\ - \mu_0 M_s (p(\mathbf{e}_z \cdot \mathbf{m}(\mathbf{x})) + 1)H_{bias} + \frac{\mu_0 M_s^2}{8\pi} \int d^3x' \frac{(\vec{\nabla}_{\mathbf{x}} \cdot \mathbf{m}(\mathbf{x}))(\vec{\nabla}_{\mathbf{x}'} \cdot \mathbf{m}(\mathbf{x}'))}{|\mathbf{x} - \mathbf{x}'|} \end{array} \right), \quad (5)$$

where $A_{ex}$ is the exchange stiffness constant of the spin-wave, $D$ is the interfacial DMI constant, and $K_u$ is the uniaxial magnetic anisotropy constant. The external magnetic bias field, $H_{bias}$, is assumed to be uniform and perpendicular to the film plane. In this paper, the bulk-type DMI is ignored.

At the equilibrium, the minimization condition of the internal energy provides the following conditions for $\theta(r)$ and the helicity $h$;

$$\begin{cases} A_{ex}\left(\frac{\partial}{\partial r} r \frac{\partial \theta}{\partial r} - \frac{\sin 2\theta}{2r}\right) - p|D|\sin^2\theta - K_{u,eff}\frac{r}{2}\sin 2\theta - \frac{\mu_0 M_s H_{bias}}{2} r\sin\theta = 0 \\ \cos(h)D = -p|D| \\ K_{u,eff} \equiv K_u - \frac{1}{2}\mu_0 M_s^2 \end{cases} \quad . \quad (6)$$

The first and second equations in (6) determine the function $\theta(r)$, although the analytical solution for $\theta(r)$ is unknown. The second equation determines the helicity of the skyrmion. Here, we are concerned only with the Néel skyrmion. The local dipole interaction is included only in $K_{u,eff}$. The treatment is valid for relatively small skyrmions (<100nm) with large interfacial DMI. In order to describe larger skyrmion bubbles, the non-local dipole interaction is essential. Details are explained in Appendix I.

Figure 1 (a) shows the magnetization distribution of a Néel skyrmion with $n_{sk}=1$, $p=1$, and $h=0$. In figure 1 (b), the cross-sectional profile of $m_z$ is shown. The radius of the skyrmion, $R$, is defined as $r$ at which $m_z = 0$. The transient region (domain wall) is defined as the region which $|m_z|$ is smaller than about 0.5. $\Delta$ expresses the width of the transient region (A more concrete definition is given in equation (17)).

As seen above, the general state of a skyrmion is determined by $n_{sk}$, $p$, $h$, $R$, $\Delta$, and the position of the skyrmion center $\mathbf{X}$. Among them, $n_{sk}$ and $p$ are integers and unchanged during the motion, while $h$, $R$, $\Delta$, and $\mathbf{X}$ can be dynamical variables. For the case of the straight domain wall, its position and azimuth angle of magnetization are canonically conjugate variables to describe its dynamics. For the case of the skyrmion, the position of the transition region is determined by $R$ and $\mathbf{X}$. Therefore, it is natural to choose $h$, $R$, and $\mathbf{X}$ as collective coordinates. For the sake of convenience, we express $R$ and $h$ into time-independent equilibrium values $R_{eq}, h_{eq}$ plus time-dependent dynamical variables $\rho(\varphi,t), \psi(\varphi,t)$,

$$\begin{cases} \theta(r,\varphi,t) = \theta(r;\tilde{R}_{eq}) + \left.\dfrac{\partial \theta(r;\tilde{R})}{\partial \tilde{R}}\right|_{\tilde{R}=\tilde{R}_{eq}} \rho(\varphi,t) \\ \phi(\varphi,t) = \varphi + h_{eq} + \psi(\varphi,t) \\ \begin{pmatrix} x \\ y \end{pmatrix} = \Delta \begin{pmatrix} \tilde{X}_x(t) \\ \tilde{X}_y(t) \end{pmatrix} + \begin{pmatrix} r\cos\varphi \\ r\sin\varphi \end{pmatrix} \end{cases} \quad , (7)$$

where $\tilde{R} \equiv R/\Delta$, and $(\tilde{X}_x, \tilde{X}_y) \equiv (X_x, X_y)/\Delta$ are the dimensionless radius and position of the skyrmion. Hereafter, variables with a tilde on top stand for dimensionless variables. In this paper, the first equation in (7) expresses a change in the radius. This definition is equivalent to $\tilde{R}(\varphi,t) = \tilde{R}_{eq} + \rho(\varphi,t)$ within the linear approximation. The function $\theta(r;\tilde{R})$ is determined to minimize the total energy of the skyrmion for a given $\tilde{R}$. When $\tilde{R} = \tilde{R}_{eq}$, the skyrmion has minimum energy and is in equilibrium. It should be noted that $\psi(\varphi,t)$ is not a function of $r$. This is one of the crucial assumptions in this paper. The polar coordinates $(r,\varphi)$ are taken as relative coordinates for $(\tilde{X}_x, \tilde{X}_y)$ to express the position inside the skyrmion. As a result, the absolute position $(x,y)$ is expressed by the 3$^{rd}$ equation in (7).

$$\begin{cases} \rho(\varphi,t) = \sum\limits_{n=0}^{+\infty} \sum\limits_{\nu=\pm 1} \rho_{n,\nu}(t) T_{n,\nu}(\varphi) \\ \psi(\varphi,t) = \sum\limits_{n=0}^{+\infty} \sum\limits_{\nu=\pm 1} \psi_{n,\nu}(t) T_{n,\nu}(\varphi) \end{cases}, \quad \begin{cases} T_{n,+1}(\varphi) = \cos n\varphi \\ T_{n,-1}(\varphi) = \sin n\varphi \end{cases} \quad (8)$$

In addition, $\rho(\varphi,t)$ and $\psi(\varphi,t)$ are expanded to the Fourier series, as shown in the above equation. The non-negative integer $n = 0,1,2,...$ is the Fourier indices corresponding to the $n$-th harmonics, $n=0$ corresponds to the breathing mode, $n=1$ parallel displacement, and *n*=2 elliptical deformation. The other subscript $\nu = \pm 1$ is parity. In the end, the collective coordinates are $\{\tilde{X}_x, \tilde{X}_y, \rho_{0,1}(t), \psi_{0,1}(t),$ $\rho_{1,1}(t), \rho_{1,-1}(t), \psi_{1,1}(t), \psi_{1,-1}(t), \rho_{2,1}(t), \rho_{2,-1}(t), \psi_{2,1}(t), \psi_{2,-1}(t)...\}$. The residual degree of the freedom that describes the extended spin wave is neglected in this paper.

The dynamics of a magnetic texture with a fixed size of vector modulus are well described by the LLG equation (see Equation (9)). The LLG equation approximates atomic magnetization by a continuous vector field and quantum transition by classical time evolution. In equation (9), all interactions are included in the "effective field" $\mathbf{H}_{\textit{eff}}$, through the energy functional $U[\mathbf{m}] = U^{\text{int}}[\mathbf{m}] + U^{\text{ext}}[\mathbf{m}]$. Here, $U^{\text{ext}}[\mathbf{m}]$ is the external energy functional that arises because of external interactions with the external magnetic field, for example. In this paper, the classical skyrmion is defined as an object that obeys the LLG equation.

$$\begin{cases} \dfrac{d\mathbf{m}}{dt} = \gamma \mathbf{m} \times \mathbf{H}_{\textit{eff}} + \alpha \mathbf{m} \times \dfrac{d\mathbf{m}}{dt} \\ \mathbf{H}_{\textit{eff}}(\mathbf{x},t;[\mathbf{m}]) = -\dfrac{1}{\mu_0 M_s} \dfrac{\delta U}{\delta \mathbf{m}} \end{cases} \quad (9)$$

The equation of motion of the collective coordinates was shown by Tretiakov et al.[30]. If a 2-dimensional magnetic texture, which obeys the LLG equation, is represented by collective coordinates $\xi(t) = \{\xi_1(t), \xi_2(t),...\}$, the collective coordinates satisfy the following set of equations.

$$\begin{cases} G_{ij}\dot{\xi}_j - \Gamma_{ij}\dot{\xi}_j + F_i^{\text{int}} + F_i^{\text{ext}} = 0 \\ G_{ij} = a\int d^2 x \mathbf{m} \cdot \left(\dfrac{\partial \mathbf{m}}{\partial \xi_i} \times \dfrac{\partial \mathbf{m}}{\partial \xi_j}\right) \\ \Gamma_{ij} = \alpha a \int d^2 x \dfrac{\partial \mathbf{m}}{\partial \xi_i} \cdot \dfrac{\partial \mathbf{m}}{\partial \xi_j} \\ F_i^{\text{int}} = -\dfrac{\partial U^{\text{int}}}{\partial \xi_i} \end{cases} \quad (10)$$

It should be noted that the theory uses magneto-static approximation [32], in which the magnetic dipole coupling is immediate. $F_i^{\text{int}}$ and $F_i^{\text{ext}}$ are an internal and an external force, respectively. The Thiele equation can be obtained from equation (10) by taking the steady-state magnetic texture position as collective coordinates.

**Linearized equation of motion**

Equation (8) is different from the Thiele equation since the coefficients $G_{ij}$ and $\Gamma_{ij}$ generally are functions of the collective coordinates. Moreover, it forms an infinite-dimensional non-linear equation system for our model. Linear approximation and decoupling are introduced to make the equation system into a block-diagonal form and tractable.

First, we consider the internal energy functional $U^{int}$. Because of the translational symmetry of the system, $U^{int}$ is given as a functional of $\rho(\varphi,t)$, $\psi(\varphi,t)$ but not of $\tilde{\mathbf{X}}^t \equiv (\tilde{X}_x(t), \tilde{X}_y(t))$. In calculating the internal force within linear approximation, only quadratic expansion of $U^{int}$ in $\rho(\varphi,t), \psi(\varphi,t)$ is necessary. Since $\rho(\varphi,t) = \psi(\varphi,t) = 0$ is a stable equilibrium in the system, the first-order expansion terms are zero, and the quadratic form must be positive-definite. By considering the non-local property of the interaction, one may expand $U^{int}$ as

$$U^{int} = U^{int}_{eq} + \int_0^{2\pi} d\varphi_1 d\varphi_2 \sum_{i,j} f_{ij}(\varphi_1, \varphi_2) \xi_i(\varphi_1) \xi_j(\varphi_2).$$

Here, $\xi_i(\varphi)$ represents $\rho(\varphi,t)$ or $\psi(\varphi,t)$. Because of the rotational symmetry of the skyrmion, $f_{ij}(\varphi_1, \varphi_2)$ is only a function of $\varphi_1 - \varphi_2$ and periodic. Therefore, it can be expanded in the Fourier series. Then, one may easily find that the quadratic form is diagonal with respect to the Fourier index $n$. Non-diagonal elements, however, may exist between $\rho_{n,\nu}(t)$ and $\psi_{n,\nu'}(\varphi)$ and causes chiral property of the bounded spin-wave.

Second, the coefficient matrices, $G_{ij}$ and $\Gamma_{ij}$, are examined. For this purpose, $\dfrac{\partial \mathbf{m}}{\partial \xi_i}$ are calculated as follows,

$$\begin{cases} \dfrac{\partial \mathbf{m}}{\partial \widetilde{X}_x} = -\Delta \dfrac{\partial \mathbf{m}}{\partial x} = -\Delta \left( T_{1,1} \dfrac{\partial \theta}{\partial r} \mathbf{e}_\theta - T_{1,-1} \dfrac{1}{r} \sin\theta \mathbf{e}_\phi \right) + O(\xi) \\ \dfrac{\partial \mathbf{m}}{\partial \widetilde{X}_y} = -\Delta \dfrac{\partial \mathbf{m}}{\partial y} = -\Delta \left( T_{1,-1} \dfrac{\partial \theta}{\partial r} \mathbf{e}_\theta + T_{1,1} \dfrac{1}{r} \sin\theta \mathbf{e}_\phi \right) + O(\xi) \\ \dfrac{\partial \mathbf{m}}{\partial \rho_{n,\nu}} = T_{n,\nu}(\varphi) \dfrac{\partial \theta}{\partial \widetilde{R}} \mathbf{e}_\theta \\ \dfrac{\partial \mathbf{m}}{\partial \psi_{n,\nu}} = T_{n,\nu}(\varphi) \sin\theta \mathbf{e}_\phi \end{cases} \quad , \quad (11)$$

where $\dfrac{\partial \mathbf{m}}{\partial \theta} = \mathbf{e}_\theta$, $\dfrac{\partial \mathbf{m}}{\partial \phi} = \sin\theta \mathbf{e}_\phi$.

In equation (11), since $\theta$ depends on $\varphi$, $G_{ij}$ and $\Gamma_{ij}$ are not diagonalized by the Fourier transformation. A linearized equation of motion will be considered to keep the problem simple. Now, $\rho_{n,\nu}(t)$ and $\psi_{n,\nu}(t)$ are taken as infinitesimally small valuables. $(\widetilde{X}_x(t), \widetilde{X}_y(t))$ are not small but $\dfrac{d}{dt}(\widetilde{X}_x(t), \widetilde{X}_y(t))$ are treated as small valuables. Under linear approximation, $G_{ij}, \Gamma_{ij}$ should be evaluated in the zeroth order of those small valuables. For this case, $\dfrac{\partial \mathbf{m}}{\partial \widetilde{X}_x}, \dfrac{\partial \mathbf{m}}{\partial \widetilde{X}_y}$ include only the $n=1$ Fourier component. And $\dfrac{\partial \mathbf{m}}{\partial \rho_{n,\nu}}, \dfrac{\partial \mathbf{m}}{\partial \psi_{n,\nu}}$ include only $n,\nu$ components. This means that $G_{ij}$ and $\Gamma_{ij}$ in this approximation are diagonal for the Fourier index. As a result, the equation of motion including $(\widetilde{X}_x(t), \widetilde{X}_y(t))$ is block diagonalized for $n=1$ as follows,

$$(\hat{G} - \hat{\Gamma}) \dfrac{d}{dt} \xi(t) - \hat{\kappa} \xi(t) + \mathbf{F}(t) = 0, \quad (12)$$

where

$\xi^t \equiv (\xi_1, \xi_2, \xi_3, \xi_4, \xi_5, \xi_6) \equiv (\widetilde{X}_x, \widetilde{X}_y, \psi_{1,1}, \psi_{1,-1}, \rho_{1,1}, \rho_{1,-1})$ and $\mathbf{F}^t \equiv (F_1, F_2, F_3, F_4, F_5, F_6)$

$\equiv (F_{\widetilde{X}_x}, F_{\widetilde{X}_y}, F_{\psi_{1,1}}, F_{\psi_{1,-1}}, F_{\rho_{1,1}}, F_{\rho_{1,-1}}) \equiv (\mathbf{F}_{\widetilde{X}}^t, \mathbf{F}_\psi^t, \mathbf{F}_\rho^t)$. Hereafter, indices $\{1,2,3,4,5,6\}$ are equivalent to $\{\widetilde{X}_x, \widetilde{X}_y, \psi_{1,1}, \psi_{1,-1}, \rho_{1,1}, \rho_{1,-1}\}$. $F_j$ is a such external force that works on the coordinate $\xi_j$ of the skyrmion. For example, the spatially modulated perpendicular magnetic field makes an energy gradient for $\tilde{\mathbf{X}}$ and $\mathbf{\rho}^t \equiv (\rho_{1,1}, \rho_{1,-1})$ but does not make it for $\mathbf{\psi}^t \equiv (\psi_{1,1}, \psi_{1,-1})$. Therefore, it produces a force acting only on $\tilde{\mathbf{X}}$ and $\mathbf{\rho}$.

$\hat{G}$ is an $6\times 6$ asymmetric matrix, whereas $\hat{\Gamma}$ and $\hat{\kappa}$ are $6\times 6$ symmetric matrices. The matrices follow rotational symmetry and selection by parity. In addition, $\hat{G}$ is an odd function of the polarization $p$. After all, out of 72 $G_{ij}$'s and $\Gamma_{ij}$'s, only 9 elements are independent. Concrete expressions of those elements are shown in equations (13) and (14).

$$\begin{cases} \dfrac{\hat{G}-\hat{\Gamma}}{2\pi a \Delta^2} = \begin{pmatrix} -\alpha(\langle \tilde{r} \rangle_{-1,-1} + \langle \tilde{r}^{-1} \rangle_{1,1}) & 2p & -p\langle \tilde{r} \rangle_{0,0} & -\alpha\langle 1 \rangle_{1,1} & -\alpha\langle \tilde{r} \rangle_{0,-1} & p\langle 1 \rangle_{1,0} \\ -2p & -\alpha(\langle \tilde{r} \rangle_{-1,-1} + \langle \tilde{r}^{-1} \rangle_{1,1}) & \alpha\langle 1 \rangle_{1,1} & -p\langle \tilde{r} \rangle_{0,0} & -p\langle 1 \rangle_{1,0} & -\alpha\langle \tilde{r} \rangle_{0,-1} \\ p\langle \tilde{r} \rangle_{0,0} & \alpha\langle 1 \rangle_{1,1} & -\alpha\langle \tilde{r} \rangle_{1,1} & 0 & p\langle \tilde{r} \rangle_{1,0} & 0 \\ -\alpha\langle 1 \rangle_{1,1} & p\langle \tilde{r} \rangle_{0,0} & 0 & -\alpha\langle \tilde{r} \rangle_{1,1} & 0 & p\langle \tilde{r} \rangle_{1,0} \\ -\alpha\langle \tilde{r} \rangle_{0,-1} & p\langle 1 \rangle_{1,0} & -p\langle \tilde{r} \rangle_{1,0} & 0 & -\alpha\langle \tilde{r} \rangle_{1,-1} & 0 \\ -p\langle 1 \rangle_{1,0} & -\alpha\langle \tilde{r} \rangle_{0,-1} & 0 & -p\langle \tilde{r} \rangle_{1,0} & 0 & -\alpha\langle \tilde{r} \rangle_{1,-1} \end{pmatrix} \\ \equiv \dfrac{1}{2\pi a \Delta^2}\begin{pmatrix} \hat{\gamma}_{\tilde{X}\tilde{X}} & \hat{\gamma}_{\tilde{X}\psi} & \hat{\gamma}_{\tilde{X}\rho} \\ \hat{\gamma}_{\psi\tilde{X}} & \hat{\gamma}_{\psi\psi} & \hat{\gamma}_{\psi\rho} \\ \hat{\gamma}_{\rho\tilde{X}} & \hat{\gamma}_{\rho\psi} & \hat{\gamma}_{\rho\rho} \end{pmatrix} \\ \hat{\kappa} = \begin{pmatrix} 0 & 0 & 0 & 0 & 0 & 0 \\ 0 & 0 & 0 & 0 & 0 & 0 \\ 0 & 0 & \kappa_{\psi\psi} & 0 & 0 & \kappa_{\psi\rho} \\ 0 & 0 & 0 & \kappa_{\psi\psi} & -\kappa_{\psi\rho} & 0 \\ 0 & 0 & 0 & -\kappa_{\psi\rho} & \kappa_{\rho\rho} & 0 \\ 0 & 0 & \kappa_{\psi\rho} & 0 & 0 & \kappa_{\rho\rho} \end{pmatrix} \end{cases},$$

(13)

where $\langle ... \rangle_n$ is the weighted average centered at the transient region (see equation (14)). $\hat{\gamma}_{ij}$ are $2\times 2$ matrices and commutative. The special structure of $\hat{\gamma}_{ij}$ validates a representation using complex numbers. The method is shown in Appendix II. Equations in (14) are evaluated at the equilibrium of the skyrmion. Apparently, $\langle 1 \rangle_{0,0} = 1$, and for large skyrmions, $\langle f(\tilde{r}) \rangle_{n,m} \approx f(\tilde{R}_{eq})$ is expected.

$$\begin{cases} \langle f(\tilde{r}) \rangle_{n,m} \equiv p\dfrac{1}{2}\int_0^\infty d\tilde{r}\, f(\tilde{r})\Omega_1(\tilde{r})^n \Omega_2(\tilde{r})^m \sin\theta \dfrac{\partial \theta}{\partial \tilde{r}} \\ \Omega_1(\tilde{r}) \equiv -\dfrac{\partial \theta}{\partial \tilde{R}}\Big/\dfrac{\partial \theta}{\partial \tilde{r}} \\ \Omega_2(\tilde{r}) \equiv -p\sin\theta\left(\dfrac{\partial \theta}{\partial \tilde{R}}\right)^{-1} \\ \kappa_{\xi_1 \xi_2} = \dfrac{\partial^2 U^{\text{int}}}{\partial \xi_1 \partial \xi_2} \end{cases} \qquad (14)$$

In the above $\kappa_{\xi_1 \xi_2}$ is the force coefficient. The expressions (13) and (14) are exact under the linear approximation if the exact profile of the skyrmion around the equilibrium is known.

It should be noted that the screening effect [22] is included in equation (9) within the linear approximation. The effective field induced by a displacement in the collective coordinates with $n \neq 1$ does not couple with $\left(\tilde{X}_x(t), \tilde{X}_y(t)\right)$ as well as $\rho_{1,1}, \rho_{1,-1}, \psi_{1,1}, \psi_{1,-1}$. Therefore, a screening does not occur through the channels with $n \neq 1$. The effective field that is made by $\rho_{1,1}, \rho_{1,-1}, \psi_{1,1}, \psi_{1,-1}$ is already considered in $U^{\text{int}}$. Therefore, it is unnecessary to discuss additional screening except for the screening by the extended spin waves.

The physically sound position of the skyrmion is not $\tilde{\mathbf{X}}$ but the center of the skyrmion charge $\tilde{\mathbf{X}}^C$ and/or center of the magnetization $\tilde{\mathbf{X}}^M$.

$$\begin{cases} \tilde{\mathbf{X}}^C = \dfrac{\int dxdy\, \tilde{\mathbf{X}}\left(\mathbf{m}\cdot\left(\dfrac{\partial \mathbf{m}}{\partial x} \times \dfrac{\partial \mathbf{m}}{\partial y}\right)\right)}{\int dxdy\left(\mathbf{m}\cdot\left(\dfrac{\partial \mathbf{m}}{\partial x} \times \dfrac{\partial \mathbf{m}}{\partial y}\right)\right)} = \tilde{\mathbf{X}} + \dfrac{1}{2}\langle \tilde{r} \rangle_{0,0} \hat{E}\boldsymbol{\psi} + \dfrac{1}{2}\langle 1 \rangle_{1,0} \boldsymbol{\rho} \\ \tilde{\mathbf{X}}^M = \dfrac{\int dxdy\, \tilde{\mathbf{X}}(\mathbf{m}\cdot\mathbf{e}_z + p)}{\int dxdy(\mathbf{m}\cdot\mathbf{e}_z + p)} = \tilde{\mathbf{X}} + \dfrac{\langle \tilde{r}^2 \rangle_{1,0}}{\langle \tilde{r}^2 \rangle_{0,0}} \boldsymbol{\rho} \end{cases} \qquad (15)$$

Here, $\hat{E} = \begin{pmatrix} 0 & 1 \\ -1 & 0 \end{pmatrix}$.

Defining $\{\tilde{\mathbf{X}}^C, \boldsymbol{\psi}, \boldsymbol{\rho}\}$ or $\{\tilde{\mathbf{X}}^M, \boldsymbol{\psi}, \boldsymbol{\rho}\}$ as new collective coordinates, equations of motions are rewritten as those presented in Appendix II. Especially for the case of zero-damping, the system of the equations for the charge center can be reduced to,

$$-G_{xy}\mathbf{e}_z \times \frac{d}{dt}\mathbf{X}^C + \mathbf{F}_X = 0. \qquad (16)$$

It is the Thiele equation for the massless particle. The result is consistent with ref. [23, 24].

In general, simultaneous differential equations must be solved to obtain $\tilde{\mathbf{X}}^C(t)$ and $\tilde{\mathbf{X}}^M(t)$ as a function of the applied external force. If one tries to remove $\{\boldsymbol{\psi}, \boldsymbol{\rho}\}$ from the equations, the differential equation for $\tilde{\mathbf{X}}^{C,M}(t)$ will include first to third-time derivatives of the variables, which will be impractical. One may also use Fourier transform to obtain the response function. If we separate the response function into the gyration term, dissipation term, and acceleration term, it will define frequency-dependent gyro-vector, dissipation dyadic, and mass similar to those introduced in ref. [22]. The separation, however, may include ambiguity. Therefore, in this section, the equation of motion (12) is treated without reducing $\{\boldsymbol{\psi}, \boldsymbol{\rho}\}$ and will be integrated numerically. From the numerical results, trajectories of the center of the skyrmion charge/magnetic moment will be calculated using equation (15).

In Fig. 2, trajectories of the skyrmion are displayed under a step-like external force that can be applied using a spatially inhomogeneous perpendicular magnetic field. In Fig. 2 (a), the unit size of the inhomogeneous external field is 1.22 MT/m, and the unit time is 57.4 ps. Fig. 2 (b) shows trajectories obtained by a micromagnetic simulation. The LLG simulation was done for the 512 nm ×512 nm ×1.2 nm ferromagnetic film using the MuMax3 simulator [34]. The film was divided into a 1 nm ×1 nm ×1.2nm mesh. For the LLG simulation following material parameters are employed. $M_s$=580 kA/m, $d$ =1.2 nm, $-\gamma/(2\pi\mu_0)$=28 GHz/T. Uniaxial magnetic anisotropy constant, $K_u$ =0.9 MJ/m$^3$, exchange constant, $A_{ex}$=27 pJ/m, interfacial DMI constant, $D$=−4.5 mJ/m, damping constant $\alpha = 0.1$, and temperature, $T$=0 K. The parameters are the same as those used in ref. 14, except for the damping constant and the temperature. The obtained skyrmion has its radius, $R_{eq}$=10.6 nm, and transition region thickness, $\Delta$= 5.3 nm at 0 K. The orange curve that presents the trajectory of the center of the charge shows a typical mass-less response of the skyrmion. In contrast, the center of the magnetization that is presented by the blue curve shows gyration and the inertial motion of the skyrmion. The inertial property is evident at the end of the trajectory.

Fig. 2 (c) shows the trajectories obtained by integrating equation (12). The same parameters employed for the LLG simulation are used for the calculation. The trajectories are essentially the same as that obtained by the micromagnetic simulation in terms of the trajectory size and appearance, although the gyration frequency is slightly different. Semi-quantitative agreement between the LLG simulation and equation (12) demonstrates that only six collective coordinates can express the essential dynamics of the skyrmion.

**Generalized Thiele equation and mass**

It was demonstrated in the previous section that the $6\times 6$ matrix expression of the equation for the skyrmion motion gives satisfactory results in comparison with LLG simulation. In this section, the $2\times 2$ matrix expression of the dynamical equation, i.e., the generalized Thiele equation, will be derived under certain conditions.

From equations (13), one can directly calculate that $\det \hat{G} = (2\pi a \Delta^2)^6 \langle \tilde{r} \rangle_{1,0}^2 \left( 2\langle \tilde{r} \rangle_{0,0} \langle 1 \rangle_{1,0} - 2\langle \tilde{r} \rangle_{1,0} \right)^2$. Therefore, a large skyrmion limit $\det \hat{G} \to 0$ is expected. This property allows us to construct a simplified equation of motion for the skyrmion. In order to estimate the size dependence of the matrix elements, Wang's expression of the skyrmion profile is utilized [27].

$$\begin{cases} \theta(r;R) = 2\tan^{-1}\left[\left(\dfrac{\sinh[\tilde{r}]}{\sinh[\tilde{R}]}\right)^p\right] \\ \Omega_1(r;R) = \dfrac{\tanh[\tilde{r}]}{\tanh[\tilde{R}]} \\ \Omega_2(r;R) = \tanh[\tilde{R}] \end{cases} \qquad (17)$$

The equation (17) is relatively accurate in a wide range of its radius ($\tilde{R}_{eq} > 0.5$). In figure 4, matrix elements evaluated using Wang's expression are plotted as a function of $\tilde{R}_{eq}$. As shown in figure 4, the approximation $\langle f(\tilde{r}) \rangle_{n,m} \approx f(\tilde{R}_{eq})$ has less than 10% error when $\tilde{R}_{eq} > 1.9$ and less than 5% error for $\tilde{R}_{eq} > 3.9$. Therefore, except for very small skyrmions ($\tilde{R}_{eq} < 2$), the approximation $\langle f(\tilde{r}) \rangle_{n,m} \approx f(\tilde{R}_{eq})$ will be semi-quantitatively accessible.

First, the approximation helps obtain skyrmion size in equilibrium, as shown by Wang *et al.*[27]. Details are in Appendix III. Second, the approximation leads to $\det(\hat{G} - \hat{\Gamma}) \cong 0$. As a result, the following relation, which does not include time derivative, is obtained;

$$(\kappa_{\psi\psi} - \tilde{R}_{eq}\kappa_{\psi\rho})\hat{E}\psi + (\tilde{R}_{eq}\kappa_{\rho\rho} - \kappa_{\psi\rho})\rho + (\tilde{R}_{eq}\mathbf{F}_{\tilde{X}} - \hat{E}\mathbf{F}_{\psi} - \tilde{R}_{eq}\mathbf{F}_{\rho}) \cong 0. \tag{18}$$

Using this relation, $\rho$ is eliminated from the equation of motion, and $4 \times 4$ expressions are obtained in Appendix IV.

Finally, $\psi$ is eliminated from the system to obtain the equations of motion of the sole $\tilde{\mathbf{X}}^{C,M}(t)$. In the derivation, the terms including $\alpha F$ and $\alpha^2$ are neglected since those terms originate in higher order terms in the force/stochastic force. For the center of charge, if the motion described by $\psi$ is rapid (much greater than the bounded spin wave frequency), the following equation of motion is obtained;

$$\begin{cases} -G_{xy}\mathbf{e}_z \times \dfrac{d}{dt}\mathbf{X}^C - \Gamma'_{xx}\dfrac{d}{dt}\mathbf{X}^C + \mathbf{F}_X = 0 \\ G_{xy} = 4\pi a n_{sk} \\ \Gamma'_{xx} = 4\pi a \tilde{R}_{eq}^{-1} \end{cases} \tag{19}$$

It should be noted that we restore the dimension of the valuables in the above equation and thereafter. The equation is the Thiele equation for the massless skyrmion. However, the dissipation dyadic is smaller than that in equation (2).

On the other hand, the acceleration term with tensorial mass appears for the low-speed case. The mass tensor, however, is off-diagonal and small because it is proportional to the Gilbert damping.

$$\begin{cases} \hat{m}^C \dfrac{d^2\mathbf{X}^C}{dt^2} = -G_{xy}\mathbf{e}_z \times \dfrac{d\mathbf{X}^C}{dt} - \Gamma_{xx}\dfrac{d\mathbf{X}^C}{dt} + \mathbf{F}_X = 0 \\ \hat{m}^C = -p\alpha(2\pi a)^2 \dfrac{(\tilde{R}_{eq} - \tilde{R}_{eq}^{-1})(R_{eq}^2\kappa_{\rho\rho} - 2\Delta R_{eq}\kappa_{\psi\rho} + \Delta^2\kappa_{\psi\psi})}{2(\kappa_{\psi\psi}\kappa_{\rho\rho} - \kappa_{\psi\rho}\kappa_{\psi\rho})}\begin{pmatrix} 0 & 1 \\ -1 & 0 \end{pmatrix} \end{cases}, \tag{20}$$

where $\hat{m}^C$ is the tensorial mass.

Finally, the center of magnetization obeys the following generalized Thiele equation.

$$\begin{cases} \hat{m}^M \dfrac{d^2 \mathbf{X}^M}{dt^2} = -G_{xy}\mathbf{e}_z \times \dfrac{d\mathbf{X}^M}{dt} - \Gamma_{xx}\dfrac{d\mathbf{X}^M}{dt} + \mathbf{F}_X + \dfrac{d}{dt}\mathbf{F}_X^M \\[2mm] \hat{m}^M \equiv (2\pi a)^2 \dfrac{R_{eq}^{\;2}\kappa_{\rho\rho} - 2\Delta R_{eq}\kappa_{\psi\rho} + \Delta^2 \kappa_{\psi\psi}}{\kappa_{\psi\psi}\kappa_{\rho\rho} - \kappa_{\psi\rho}\kappa_{\psi\rho}} \begin{pmatrix}1 & 0 \\ 0 & 1\end{pmatrix} \\[2mm] \mathbf{F}_X^M \equiv -\dfrac{1}{\Delta}\dfrac{1}{\kappa_{\psi\psi}\kappa_{\rho\rho} - \kappa_{\psi\rho}\kappa_{\psi\rho}} \left( \begin{pmatrix}0 & 1 \\ -1 & 0\end{pmatrix}(\mathbf{F}_{\tilde{X}} - \mathbf{F}_\rho)(\tilde{R}_{eq}\kappa_{\psi\rho} - \kappa_{\psi\psi}) + p\mathbf{F}_\psi(\tilde{R}_{eq}\kappa_{\rho\rho} - \kappa_{\psi\rho}) \right) \end{cases}$$

(21)

where $m^M$ is the scalar mass of the center of the magnetization. The equation includes acceleration term with scalar mass and time derivative of the external force [34]. The skyrmion shows inertia because it acquires the energy in the internal degree of freedom. While the term $\dfrac{d}{dt}\mathbf{F}_X^M$ facilitates the immediate response of skyrmion. The massless feature remains even though the center of the magnetization gets inertial mass. The size of the mass for the center of the magnetization is in the same order of magnitude as the Döring mass of the straight domain wall [35].

**Conclusion**

The dynamics of the magnetic skyrmion described by the LLG equation are reasonably reproduced by a set of linearized equations of motion to the collective coordinates. The coordinates of the skyrmion center and the radial and azimuthal coordinates for the magnetization fluctuations are chosen as the collective coordinates. After Fourier-transforming the fluctuations, the system of the equations is block diagonalized, and the equations of motion, including 6-collective coordinates, are obtained.

From the comparison with the micromagnetic simulation based on the LLG equation, it is concluded that the equations of motion, including 6-collective coordinates, semi-quantitatively describe the dynamics of the center of skyrmion charge and the center of the magnetization. The center of the skyrmion charge shows essentially massless properties, whereas the center of the magnetization shows both inertial and massless properties.

For a large skyrmion (the radius is reasonably larger than the transient region width), the system of the equation is reduced to a simple description of the dynamics for the charge or magnetization centers of the skyrmion. The equation of motion for the center of the skyrmion charge does not possess an inertia term (scaler mass) for the fast dynamics but an off-diagonal tensorial mass for the slow dynamics, which is proportional to the

Gilbert damping factor. In contrast, in the equation of motion for the magnetization center, a finite diagonal (scaler) mass appears, comparable to the Döring mass for the straight domain wall. The time derivatives of the external forces introduce the other massless behavior to the magnetization-center dynamics.

Acknowledgments

The authors thank Dr. G. Tatara of Riken. This work was supported by JSPS Grant-in-Aid for Scientific Research (S) Grant Number JP20H05666, Japan and CREST (Non-classical Spin project, JPMJCR20C1) of the Japan Science and Technology Agency.

Appendix I: Internal energy and energy minimization condition

The internal energy of the ultrathin disk-shape skyrmion ($d \ll R_{eq}$) shown in equation (5) in the main text is rewritten using cylindrical coordinates $(r, \varphi, z)$ as follows;

$$\begin{cases} U^{ex} = 2\pi A_{ex} d \int_0^\infty r dr \left( \left(\frac{\partial \theta}{\partial r}\right)^2 + \left(\frac{\sin \theta}{r}\right)^2 \right) \\ U^{DMI} = 2\pi D d \cos(h) \int_0^\infty r dr \left( \frac{\partial \theta}{\partial r} + \frac{\sin 2\theta}{2r} \right) \\ U^{ani} = 2\pi K_u d \int_0^\infty r dr \sin^2 \theta \\ U^{Zeeman} = -2\pi \mu_0 M_s H_{bias} d \int_0^\infty r dr (\cos \theta + p) \\ U^{dipole} = U^{surface} + U^{volume} \\ U^{surface} = -2\mu_0 M_s^2 \int_0^\infty dr\, dr' (1 - \cos \theta(r) \cos \theta(r')) (F(r, r', 0) - F(r, r', d)) \\ U^{volume} = d^2 \mu_0 M_s^2 \cos^2(h) \int dr\, dr' \rho_1(r) \rho_1(r') F(r, r', 0) \\ \rho_1(r) \equiv -\left(\frac{\partial}{\partial r} + \frac{1}{r}\right) \sin \theta \\ F(r, r', d) \equiv \frac{rr'}{\sqrt{(r+r')^2 + d^2}} K\left(\sqrt{\frac{4rr'}{(r+r')^2 + d^2}}\right) \\ K(k) \equiv \int_0^{\pi/2} dx \frac{1}{\sqrt{1 - k^2 \sin^2 x}} \end{cases} \quad (A1)$$

where $K(k)$ is the complete elliptic integral of the first kind. The coordinates $(\varphi, z)$ have been integrated out, assuming a small film thickness, in the equations (A1).

Minimization of the internal energy determines conditions for the helicity, $h$, and the functional shape of $\theta(r)$ as follows;

$$\begin{cases}
\left[A_{ex}\left(\dfrac{\partial}{\partial r}r\dfrac{\partial \theta}{\partial r}-\dfrac{\sin 2\theta}{2r}\right)+D\cos(h)\sin^2\theta - K_u\dfrac{r}{2}\sin 2\theta - \dfrac{\mu_0 M_s H_{bias}}{2}r\sin\theta\right.\\
\left.+\dfrac{1}{\pi A_{ex}}\int_0^\infty dr'\left(\begin{array}{l}\dfrac{1}{d}\sin\theta(r)\cos\theta(r')(F(r,r',0)-F(r,r',d))\\+\dfrac{d}{2}\cos^2(h)\cos\theta\left(-\dfrac{\partial}{\partial r}+\dfrac{1}{r}\right)\rho_1(r')F(r,r',0)\end{array}\right)\right]=0\\
\sin(h)(D+p\cos(h)D_{th})=0\\
\mathrm{sgn}[\cos(h)]=-p\,\mathrm{sgn}[D]\\
D_{th}\equiv\dfrac{d^2\int_0^\infty dr\int_0^\infty dr'\rho_1(r)\rho_1(r')F(r,r',0)}{p\pi d\int_0^\infty rdr\left(\dfrac{\partial\theta}{\partial r}+\dfrac{\sin 2\theta}{2r}\right)}>0
\end{cases}\quad\text{, (A2)}$$

where sgn[..] is the sign function. In the first equation of (A2), the condition for $\theta(r)$ is expressed by a complex differential equation. For relatively small skyrmions (<100 nm, for example) in an ultrathin magnetic film, dipole interaction can be approximated by a local demagnetization field perpendicular to the film plane. This approximation simplifies equation (6). However, an analytical solution is still unknown.

Besides the difficulty mentioned above, the helicity $h$ can be determined from equation (A2) as follows [31];

$$\begin{cases}
\text{Neel skyrmion}: D_{th}<|D|,\ h=\begin{cases}0:pD<0\\ \pi:pD>0\end{cases},\ D_{eff}=|D|-\dfrac{1}{2}D_{th}\\
\text{Bloch skyrmion}: D_{th}>|D|,\ \cos h=-p\dfrac{D}{D_{th}},\ D_{eff}=\dfrac{D^2}{2D_{th}}
\end{cases}\quad\text{(A3)}$$

$|D|=D_{th}$ defines the phase boundary between the Néel skyrmion and Bloch skyrmion.

For large skyrmions, $D_{th}$ coincides with that in the straight domain wall [31]. The phase boundary is determined by a competition between the interfacial DMI energy and the demagnetization energy of the transient region of the skyrmion. In the above treatment, bulk-type DMI, which stabilizes Bloch-type skyrmion, is neglected. The inclusion of the bulk type DMI is straightforward. In the main text, only the Néel type skyrmions are treated, i.e. $D_{th}<|D|$.

Appendix II Complex number representation of the equation of motion

$\hat{\gamma}_{ij}$ can be written as $\hat{\gamma}_{ij} = A_{ij}\hat{I} + B_{ij}\hat{E}$, where $\hat{I} = \begin{pmatrix} 1 & 0 \\ 0 & 1 \end{pmatrix}$ and $\hat{E} = \begin{pmatrix} 0 & 1 \\ -1 & 0 \end{pmatrix}$. The algebra among $\hat{\gamma}_{ij}$ and $\mathbf{X}, \mathbf{\psi}, \mathbf{\rho}$ is the same as that of complex numbers. Therefore, one may use the following complex representation.

$$\begin{cases} \hat{\gamma}_{ij} = A_{ij}\hat{I} + B_{ij}\hat{E} \Rightarrow \gamma_{ij} \equiv A_{ij} + iB_{ij} \\ \begin{pmatrix} \xi_i \\ \xi_{i+1} \end{pmatrix} \Rightarrow \xi \equiv \xi_i - i\xi_{i+1} \end{cases} \quad \text{(A4)}$$

The representation reduces the $6\times 6$ representation to that of a complex $3\times 3$ matrix representation as follows;

$$\begin{cases} \hat{G} = -\hat{G}^\dagger = 2\pi a \Delta^2 p \begin{pmatrix} 2i & -\langle\tilde{r}\rangle_{0,0} & i\langle 1\rangle_{1,0} \\ \langle\tilde{r}\rangle_{0,0} & 0 & \langle\tilde{r}\rangle_{1,0} \\ i\langle 1\rangle_{1,0} & -\langle\tilde{r}\rangle_{1,0} & 0 \end{pmatrix} \\ \hat{\Gamma} = \hat{\Gamma}^\dagger = 2\pi a \Delta^2 \alpha \begin{pmatrix} \langle\tilde{r}\rangle_{-1,-1} + \langle\tilde{r}^{-1}\rangle_{1,1} & i\langle\tilde{r}\rangle_{1,1} & \langle\tilde{r}\rangle_{0,-1} \\ -i\langle 1\rangle_{1,1} & \langle\tilde{r}\rangle_{1,1} & 0 \\ \langle\tilde{r}\rangle_{0,-1} & 0 & \langle\tilde{r}\rangle_{1,-1} \end{pmatrix}. \quad \text{(A5)} \\ \hat{G} - \hat{\Gamma} \equiv \begin{pmatrix} \gamma_{XX} & \gamma_{X\psi} & \gamma_{X\rho} \\ \gamma_{\psi X} & \gamma_{\psi\psi} & \gamma_{\psi\rho} \\ \gamma_{\rho X} & \gamma_{\rho\psi} & \gamma_{\rho\rho} \end{pmatrix} \\ \hat{\kappa} = \hat{\kappa}^\dagger = \begin{pmatrix} \kappa_{XX} & 0 & 0 \\ 0 & \kappa_{\psi\psi} & i\kappa_{\psi\rho} \\ 0 & -i\kappa_{\psi\rho} & \kappa_{\rho\rho} \end{pmatrix} \end{cases}$$

Appendix III: Equation of motion of the center of charge and magnetization

Equation (12), which is the equation of motion of $\{\tilde{\mathbf{X}}, \mathbf{\psi}, \mathbf{\rho}\}$, can be rearranged to that for $\{\tilde{\mathbf{X}}^C, \mathbf{\psi}, \mathbf{\rho}\}$ and $\{\tilde{\mathbf{X}}^M, \mathbf{\psi}, \mathbf{\rho}\}$ using the complex representation as follows;

$$\begin{pmatrix} \gamma_{\tilde{X}\tilde{X}} & \gamma_{\tilde{X}\psi} - i\frac{1}{2}\langle\tilde{r}\rangle_{0,0}\gamma_{\tilde{X}\tilde{X}} & \gamma_{\tilde{X}\rho} - \frac{1}{2}\langle 1\rangle_{1,0}\gamma_{\tilde{X}\tilde{X}} \\ \gamma_{\psi\tilde{X}} & \gamma_{\psi\psi} - i\frac{1}{2}\langle\tilde{r}\rangle_{0,0}\gamma_{\psi\tilde{X}} & \gamma_{\psi\rho} - \frac{1}{2}\langle 1\rangle_{1,0}\gamma_{\psi\tilde{X}} \\ \gamma_{\rho\tilde{X}} & \gamma_{\rho\psi} - i\frac{1}{2}\langle\tilde{r}\rangle_{0,0}\gamma_{\rho\tilde{X}} & \gamma_{\rho\rho} - \frac{1}{2}\langle 1\rangle_{1,0}\gamma_{\rho\tilde{X}} \end{pmatrix} \frac{d}{dt}\begin{pmatrix} \tilde{X}^C \\ \psi \\ \rho \end{pmatrix} - \begin{pmatrix} 0 & 0 & 0 \\ 0 & \kappa_{\psi\psi} & i\kappa_{\psi\rho} \\ 0 & -i\kappa_{\psi\rho} & \kappa_{\rho\rho} \end{pmatrix}\begin{pmatrix} \tilde{X}^C \\ \psi \\ \rho \end{pmatrix} + \begin{pmatrix} F_{\tilde{X}} \\ F_\psi \\ F_\rho \end{pmatrix} = 0, \quad \text{(A6)}$$

$$\begin{pmatrix} \gamma_{\tilde{X}\tilde{X}} & \gamma_{\tilde{X}\psi} & \gamma_{\tilde{X}\rho} - \frac{1}{2}\frac{\langle \tilde{r}^2 \rangle_{1,0}}{\langle \tilde{r}^2 \rangle_{0,0}}\gamma_{\tilde{X}\tilde{X}} \\ \gamma_{\psi\tilde{X}} & \gamma_{\psi\psi} & \gamma_{\psi\rho} - \frac{1}{2}\frac{\langle \tilde{r}^2 \rangle_{1,0}}{\langle \tilde{r}^2 \rangle_{0,0}}\gamma_{\psi\tilde{X}} \\ \gamma_{\rho\tilde{X}} & \gamma_{\rho\psi} & \gamma_{\rho\rho} - \frac{1}{2}\frac{\langle \tilde{r}^2 \rangle_{1,0}}{\langle \tilde{r}^2 \rangle_{0,0}}\gamma_{\rho\tilde{X}} \end{pmatrix} \frac{d}{dt}\begin{pmatrix} \tilde{X}^M \\ \psi \\ \rho \end{pmatrix} - \begin{pmatrix} 0 & 0 & 0 \\ 0 & \kappa_{\psi\psi} & i\kappa_{\psi\rho} \\ 0 & -i\kappa_{\psi\rho} & \kappa_{\rho\rho} \end{pmatrix}\begin{pmatrix} \tilde{X}^M \\ \psi \\ \rho \end{pmatrix} + \begin{pmatrix} F_{\tilde{X}} \\ F_\psi \\ F_\rho \end{pmatrix} = 0 . \quad (A7)$$

Appendix IV: Skyrmion size in equilibrium

Here, Wang's treatment of skyrmion size [27] is extended to a case in which dipole-dipole interaction is important. The full contents of equation (22) in the main text are written as follows [27-29,31];

$$U^{int} = U^{ex} + U^{DMI} + U^{ani} + U^{Zeeman} + U^{Surface} + U^{Volume}$$
$$= 4\pi d \begin{bmatrix} A_{ex}(f_1(\tilde{R}) + f_2(\tilde{R})) + pD\cosh(-f_3(\tilde{R}) - f_4(\tilde{R}))\Delta \\ + \left(K_u f_5(\tilde{R}) - p\mu_0 M_s H_z f_6(\tilde{R}) - \frac{\mu_0 M_s^2}{2\pi}f_7(\tilde{R})\right)\Delta^2 + \frac{d\mu_0 M_s^2}{2\pi}\left(-f_8(\tilde{R}) + \frac{\cos^2 h}{2}f_9(\tilde{R})\right)\Delta \end{bmatrix}, \quad (A8)$$

where $\tilde{R} = R/\Delta$ is the dimensionless radius of the skyrmion. The functions $f_1(\tilde{R}), f_2(\tilde{R}),\ldots f_6(\tilde{R})$ are the same as those defined in ref. 27. Those are also expressed using the weighted average defined in the main text, i.e. $f_1(\tilde{R}) = \langle \tilde{r} \rangle_{-1,-1}$, $f_2(\tilde{R}) = \langle \tilde{r}^{-1} \rangle_{1,1}$, $f_3(\tilde{R}) = -\left\langle \frac{\tilde{r}}{\sin\theta} \right\rangle_{0,0}$, $f_4(\tilde{R}) = -p\langle \cot\theta \rangle_{1,1}$,

$f_5(\tilde{R}) = \langle \tilde{r} \rangle_{1,1}$, $f_6(\tilde{R}) = \frac{1}{2}\langle \tilde{r}^2 \rangle_{0,0}$. $f_7(\tilde{R}), f_8(\tilde{R}), f_9(\tilde{R})$ are newly introduced functions to express dipole interaction. The definitions are as follows;

$$\begin{cases} \frac{d}{\Delta}f_7 + \left(\frac{d}{\Delta}\right)^2 f_8 = \frac{1}{\Delta^3}\int_0^\infty dr\,dr'(1-\cos\theta\cos\theta')(F(r,r',0) - F(r,r',d)) \\ f_9 = \frac{1}{\Delta}\int_0^\infty dr\,dr'\rho_1(r)\rho_1(r')F(r,r',0) \end{cases} \quad (A9)$$

where $F$ is defined in Appendix I. Here, the surface contribution is separated into two terms $f_7(\tilde{R}), f_8(\tilde{R})$, according to the different thickness dependencies. $f_9(\tilde{R})$ is the bulk contribution. Alternating asymptotic form for large skyrmion [27, 29] in (A8), we get,

$$U^{int} \cong 4\pi d\left[A_{ex}\left(\frac{R}{\Delta}+\frac{\Delta}{R}\right)-\left(\frac{\pi}{2}D_{eff}+\frac{d\mu_0 M_s^2}{2\pi}\log\left[\frac{8}{e^{1-\gamma_E}\pi}\frac{R}{\Delta}\right]\right)\frac{R}{\Delta}\Delta+\left(K_{eff}\frac{R}{\Delta}-p\frac{\mu_0 M_s H_z}{2}\left(\left(\frac{R}{\Delta}\right)^2+\frac{\pi^2}{12}\right)\right)\Delta^2\right], (A10)$$

where $\gamma_E \cong 0.577$ is the Euler's constant.

Under zero magnetic field, if the film is thin enough to neglect non-local dipole interaction, minimization in energy reproduces Wang's results [27]. Including dipole energy in the transient region, we get;

$$\begin{cases}\Delta \cong \dfrac{\pi|D_{eff}|}{4K_{eff}} \\ R_{eq} \cong \Delta\left(1-\dfrac{K_{eff}}{A_{ex}}\Delta^2\right)^{-\frac{1}{2}}\end{cases}. \quad (A11)$$

Here, the DMI energy constant is replaced by $D_{eff}$.

For large skyrmion in which non-local dipole interaction is important, i.e., "skyrmion bubble, the domain wall thickness $\Delta$ is independent of the radius and approximated as follows [31];

$$\Delta \cong \sqrt{\frac{A_{ex}}{K_{eff}} - \frac{t\mu_0 M_s^2}{4\pi K_{eff}}}. \quad (A12)$$

Then, minimization of the energy provides the following equation for the radius [29 supplemental material],

$$\begin{cases}\dfrac{\partial U^{int}}{\partial R} \cong \left(\sigma_w - \sigma_M \log\dfrac{R}{\Delta} + \sigma_H \dfrac{R}{\Delta}\right)2\pi d = 0 \\ \sigma_w \equiv 2\left(\dfrac{A_{ex}}{\Delta}+K_{eff}\Delta-\dfrac{\pi}{2}|D_{eff}|\right)-\sigma_M \log\left[\dfrac{8}{e^{-\gamma_E}\pi}\right] \\ \sigma_M \equiv \dfrac{d\mu_0 M_s^2}{\pi}, \quad \sigma_H \equiv 2\mu_0 M_s H\Delta\end{cases} \quad (A13)$$

In equation (A13), the term that is proportional to $\Delta/R_{eq} \ll 1$ is neglected. The above equation has a stable solution, if $\dfrac{\sigma_w}{\sigma_M}+\ln\dfrac{\sigma_H}{\sigma_M} \leq -1$. The solution is expressed by using the Lambert $W$ function [37] as follows;

$$R_{eq} = -\Delta\frac{\sigma_M}{\sigma_H}W_{-1}\left[-\exp\left[\frac{\sigma_w}{\sigma_M}+\ln\frac{\sigma_H}{\sigma_M}\right]\right]. \quad (A14)$$

The function is available in Mathematica by the name of *ProductLog*. The function has its maximum ($W_{-1,\max} = -1$) at $\dfrac{\sigma_w}{\sigma_M} + \ln\dfrac{\sigma_H}{\sigma_M} = -1$. As a result, the minimum size that can be described by equation (A14) is, $R_{eq,\min} = \Delta\sigma_M / \sigma_H$.

Appendix V: Generalized Thiele equation

Using equation (23) in the main text, $\rho$ is eliminated from the equation of motions (A6) and (A7). The result is as follows;

$$\begin{pmatrix} \gamma_{\tilde{X}\tilde{X}} & \gamma''^{C,M}_{\tilde{X}\psi} \\ \gamma_{\psi\tilde{X}} & \gamma''^{C,M}_{\psi\psi} \end{pmatrix} \frac{d}{dt}\begin{pmatrix} \tilde{X}^{C,M} \\ \psi \end{pmatrix} - \begin{pmatrix} 0 & 0 \\ 0 & \kappa'_{\psi\psi} \end{pmatrix}\begin{pmatrix} \tilde{X}^{C,M} \\ \psi \end{pmatrix} + \begin{pmatrix} F'^{C,M}_{\tilde{X}} \\ F'^{C,M}_{\psi} \end{pmatrix} = 0, \qquad (A15)$$

where,

$$\begin{cases} \rho \equiv \rho_\psi i\psi + F'_\rho \\ \rho_\psi \equiv \dfrac{\tilde{R}_{eq}\kappa_{\psi\rho} - \kappa_{\psi\psi}}{\tilde{R}_{eq}\kappa_{\rho\rho} - \kappa_{\psi\rho}} \\ F'_\rho \equiv -\dfrac{\tilde{R}_{eq}F_{\tilde{X}} - iF_\psi - \tilde{R}_{eq}F_\rho}{\tilde{R}_{eq}\kappa_{\rho\rho} - \kappa_{\psi\rho}} \\ \gamma'^C_{j\psi} \equiv \gamma_{j\psi} - \dfrac{i}{2}\tilde{R}_{eq}\gamma_{jX},\ \gamma'^M_{j\psi} \equiv \gamma_{j\psi},\ (j=\tilde{X},\psi,\rho) \\ \gamma'^C_{j\rho} \equiv \gamma_{j\rho} - \dfrac{1}{2}\gamma_{j\tilde{X}},\ \gamma'^M_{j\rho} \equiv \gamma_{j\rho} - \gamma_{j\tilde{X}},\ (j=\tilde{X},\psi,\rho) \\ \gamma''^{C,M}_{j\psi} \equiv \gamma'^{C,M}_{j\psi} + \gamma'^{C,M}_{j\rho} i\rho_\psi \\ \kappa'_{\psi\psi} \equiv \kappa_{\psi\psi} - \kappa_{\psi\rho}\rho_\psi \\ \kappa'_{\psi\rho} \equiv \kappa_{\psi\rho} - \kappa_{\rho\rho}\rho_\psi \\ \begin{pmatrix} F'^{C,M}_{\tilde{X}} \\ F'^{C,M}_\psi \end{pmatrix} = \begin{pmatrix} F_{\tilde{X}} \\ F_\psi \end{pmatrix} - \begin{pmatrix} 0 \\ i\kappa_{\psi\rho} \end{pmatrix}F'_\rho + \dfrac{d}{dt}\begin{pmatrix} \gamma'^{C,M}_{\tilde{X}\rho} \\ \gamma'^{C,M}_{\psi\rho} \end{pmatrix}F'_\rho \end{cases} \qquad (A16)$$

For the center of the charge, if $\psi$ is changing much higher frequency than the bounded spin wave frequency, i.e. $\left|\gamma''^C_{\psi\psi}\dfrac{d\psi}{dt}\right| \gg \left|\kappa'_{\psi\psi}\psi\right|$, we may neglect $\kappa'_{\psi\psi}$ in A15 and obtain the following equation of motion.

$$2\pi a\Delta^2(2ip - 2\alpha\tilde{R}_{eq}^{-1})\frac{d}{dt}\tilde{X}^C + F_{\tilde{X}} = 0. \qquad (A17)$$

Here, terms with $\alpha^2$, $\alpha F_j$ are neglected since those terms are higher order terms with the force/stochastic force. To derive (A17), one may use the fact that $\gamma'^C_{\tilde{X}\psi}, \gamma'^C_{\tilde{X}\rho}, \gamma''^C_{\tilde{X}\psi}$ are proportional to $\alpha$. The skyrmion follows the zero-mass Thiele equation with weaker damping.

If $\psi$ is changing slowly, with a much slower speed than the bounded spin wave frequency, i.e. $\left|\gamma''^C_{\psi\psi}\frac{d\psi}{dt}\right| \ll \left|\kappa'_{\psi\psi}\psi\right|$, we may neglect $\frac{d\psi}{dt}$ in A15 and obtain the following equation of motion.

$$\begin{cases} \tilde{m}^C \frac{d^2 \tilde{X}^C}{dt^2} = \gamma_{\tilde{X}\tilde{X}} \frac{d\tilde{X}^C}{dt} + F_{\tilde{X}} \\ \tilde{m}^C = -\frac{\gamma''_{\tilde{X}\psi}\gamma_{\psi\tilde{X}}}{\kappa'_{\psi\psi}} = (2\pi a \Delta^2)^2 \frac{-ip\alpha(\tilde{R}_{eq} - \tilde{R}_{eq}^{-1})(\tilde{R}_{eq}^2 \kappa_{\rho\rho} - 2\tilde{R}_{eq}\kappa_{\psi\rho} + \kappa_{\psi\psi})}{2(\kappa_{\psi\psi}\kappa_{\rho\rho} - \kappa_{\psi\rho}\kappa_{\psi\rho})} \end{cases}. \quad (A18)$$

Here, the acceleration term with tensorial mass appeared. However, the mass is off-diagonal and proportional to the damping factor $\alpha$.

For the center of the magnetization, the following equation of motion is obtained, considering that $\gamma''^M_{\psi\psi}$ is proportional to $\alpha$.

$$\begin{cases} \tilde{m}^M \frac{d^2 \tilde{X}^M}{dt^2} = \gamma_{\tilde{X}\tilde{X}} \frac{d\tilde{X}^M}{dt} + F_{\tilde{X}} + \frac{d}{dt}F^M_{\tilde{X}} \\ \tilde{m}^M \equiv \frac{\gamma''_{\psi\psi}\gamma_{\tilde{X}\tilde{X}} - \gamma''_{\tilde{X}\psi}\gamma_{\psi\tilde{X}}}{\kappa'_{\psi\psi}} = (2\pi a \Delta^2)^2 \frac{\tilde{R}_{eq}^2 \kappa_{\rho\rho} - 2\tilde{R}_{eq}\kappa_{\psi\rho} + \kappa_{\psi\psi}}{\kappa_{\psi\psi}\kappa_{\rho\rho} - \kappa_{\psi\rho}\kappa_{\psi\rho}} \\ F^M_{\tilde{X}} = -\frac{i(F_{\tilde{X}} - F_\rho)(\tilde{R}_{eq}\kappa_{\psi\rho} - \kappa_{\psi\psi}) + pF_\psi(\tilde{R}_{eq}\kappa_{\rho\rho} - \kappa_{\psi\rho})}{\kappa_{\psi\psi}\kappa_{\rho\rho} - \kappa_{\psi\rho}\kappa_{\psi\rho}} \end{cases}. \quad (A19)$$

**Figure captions**

　Fig. 1 Structure and coordinates of the skyrmion (a) Bird view of the Néel skyrmion. (b) The cross-sectional profile of the skyrmion. Skyrmion radius $R$ is a radius at which $m_z = 0$. The transient region is defined as the region where $|m_z|$ is less than about 0.5. And $\Delta$ is the width of the transient region (see equation (17)). (c) Polar coordinates of the magnetization vector. (d) Collective coordinates of a skyrmion.

Fig. 2 Response of the skyrmion to the external force. (a) A sequence of the external force. The time interval is 57.4 ps. The external field gradient is 1.22 MT/m. (b) Trajectories of the center of the skyrmion charge (orange) and the center of the magnetization (blue) obtained by a micromagnetic simulation. (c) Trajectories of the center of the skyrmion charge (blue) and the center of the magnetization (orange) obtained by integrating equation (10).

Fig. 3 $\tilde{R}_{eq}$ dependence of the matrix elements. (a) Elements in $\hat{G}$. (b) Elements in $\hat{\Gamma}$.

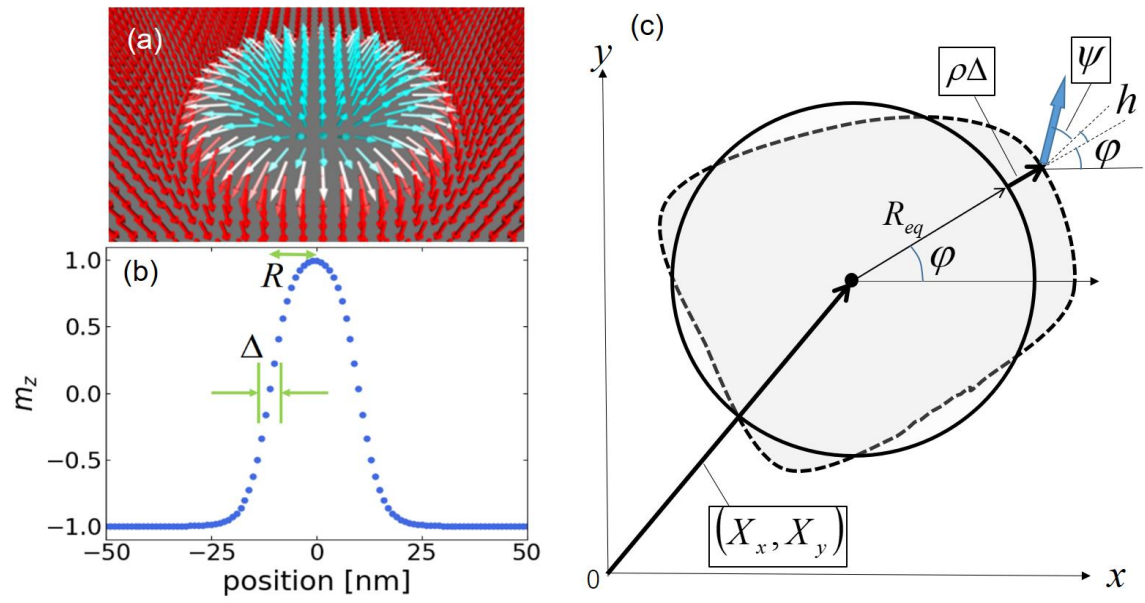

Figure 1.

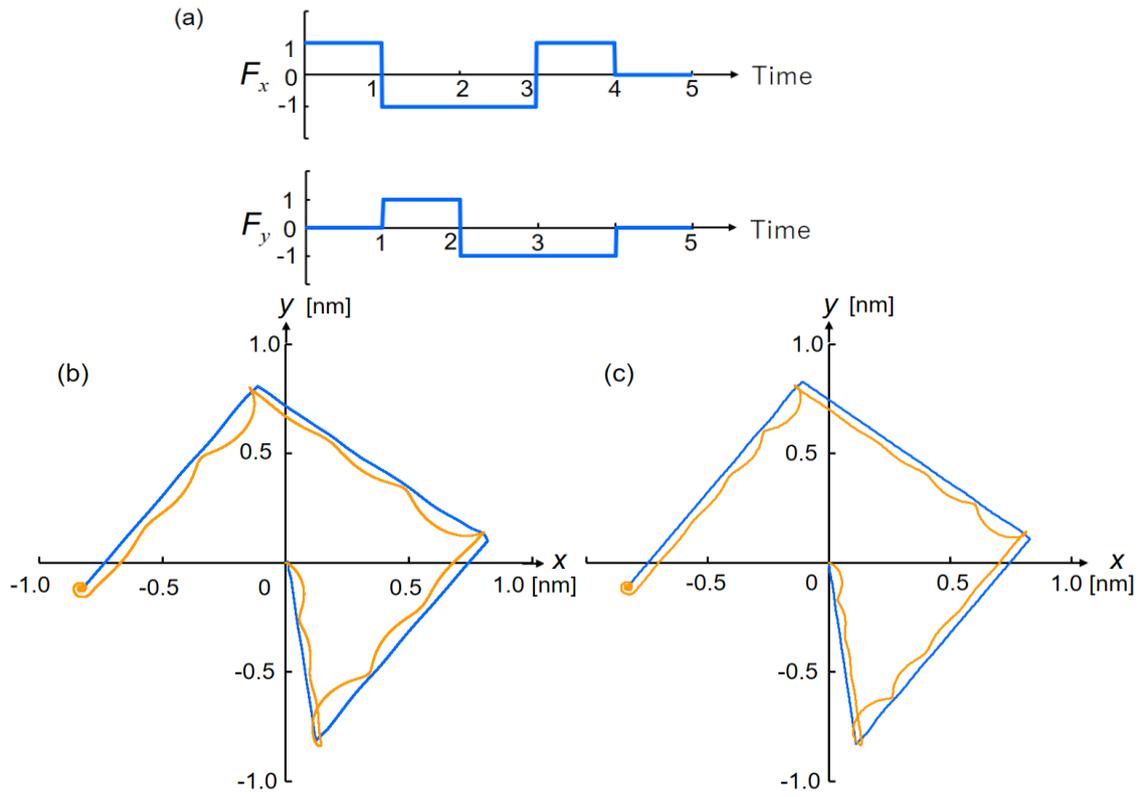

Figure 2.

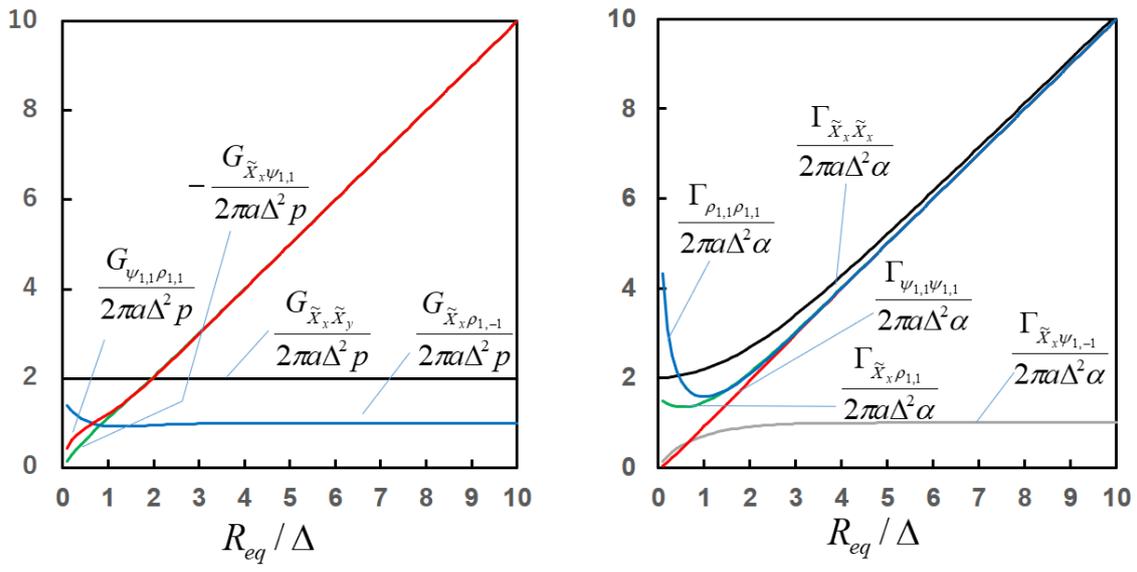

Figure 3